\documentclass{elsart}
\usepackage{graphicx,amssymb}

\def\g{\gamma}
\def\ha{\frac{1}{2}}

\def\ub#1{\underline{#1}}

\def\ran{\rangle}

\def\ket#1{|#1\ran}
\def\ha{{1\over 2}}
\def\e{\epsilon}

\newcommand{\p}{\bot}
\newcommand{\dd}{\partial}
\newcommand{\de}{\delta}
\newcommand{\ep}{\varepsilon}
\newcommand{\f}{\varphi}
\newcommand{\ls}{\left(}
\newcommand{\rs}{\right)}
\newcommand{\al}{\alpha}
\newcommand{\n}{\nu}
\newcommand{\m}{\mu}
\newcommand{\La}{\Lambda}
\newcommand{\la}{\lambda}
\newcommand{\z}{\zeta}

\newcommand{\ps}{\psi}
\newcommand{\si}{\sigma}

\begin{document}

\begin{flushright}
SLAC-PUB-10506 \\
UMN-D-04-1 \\
SMUHEP/03-13
\end{flushright}

\begin{frontmatter}

\title{A nonperturbative calculation of the electron's magnetic
moment\thanksref{thanks1}}
\thanks[thanks1]{Work supported in part by the Department of Energy
under contract numbers DE-AC03-76SF00515, DE-FG02-98ER41087, and
DE-FG03-95ER40908.}

\author[SLAC]{S. J. Brodsky,}
\author[StPete]{V. A. Franke,}
\author[UMD]{J. R. Hiller,}
\author[SMU]{G. McCartor,}
\author[StPete]{S. A. Paston,} and
\author[StPete]{E. V. Prokhvatilov}

\address[SLAC]{Stanford Linear Accelerator Center, Stanford University,
Stanford, California 94309}
\address[StPete]{St.~Petersburg State University
St. Petersburg, Russia}
\address[UMD]{Department of Physics, University of Minnesota-Duluth,
Duluth, Minnesota 55812}
\address[SMU]{Department of Physics, Southern Methodist University,
Dallas, TX 75275}

\begin{abstract}
In principle, the complete spectrum and bound-state wave functions
of a quantum field theory can be determined by finding the eigenvalues 
and eigensolutions of its light-cone Hamiltonian.  One of the challenges 
in obtaining nonperturbative solutions for gauge theories such as QCD 
using light-cone Hamiltonian methods is to renormalize the theory while 
preserving Lorentz symmetries and gauge invariance. For example, the 
truncation of the light-cone Fock space leads to uncompensated ultraviolet 
divergences.  We present two methods for consistently regularizing 
light-cone-quantized gauge theories in Feynman and light-cone gauges: 
(1) the introduction of a spectrum of Pauli--Villars fields which 
produces a finite theory while preserving Lorentz invariance; 
(2) the augmentation of the gauge-theory Lagrangian with higher derivatives.
In the latter case, which is applicable to light-cone gauge 
($A^+ = 0$), the $A^-$ component of the gauge field is maintained 
as an independent degree of freedom rather than a constraint.
Finite-mass Pauli--Villars regulators can also be used to compensate 
for neglected higher Fock states.  As a test case, we apply these 
regularization procedures to an approximate nonperturbative computation 
of the anomalous magnetic moment of the electron in QED as a first attempt 
to meet Feynman's famous challenge.
\end{abstract}


%

\end{frontmatter}

\section{Introduction}
\label{sec:Introduction}

The gyromagnetic ratio of the electron
$g_{e^-} = 2.0023193043768(86) $,
the ratio of the spin precession frequency
to the Larmor precession frequency in a static magnetic field, is an
intrinsic property of an individual lepton.  It is now known
experimentally to $12$ significant figures~\cite{VanDyck:ay} --  the
most precisely known fundamental physical parameter.  The anomalous
moment $a_e = {g_e-2\over 2}$, the deviation of the gyromagnetic
ratio  from Dirac's value $g_e=2$ due to quantum fluctuations, has
now been evaluated through order ${\alpha}^4$ in perturbative
quantum electrodynamics~\cite{Kinoshita:2002ns,Hughes:fp}.

At the 12th Solvay Conference, Feynman presented
a challenge~\cite{Feynman}: ``Is there any
method of computing the anomalous moment of the electron which, on first
approximation, gives a fair approximation to the $\alpha$ term and a
crude one to
$\alpha^2$; and when improved, increases the accuracy of the
$\alpha^2$ term, yielding a rough estimate to $\alpha^3$ and
beyond."  An interesting attempt to answer Feynman's  challenge using
sidewise dispersion relations was
pioneered by Drell and Pagels~\cite{Drell:1965hg}, but it is difficult
to make this method systematic.

The anomalous moment of a spin-half particle can be evaluated
without approximation from the overlap of its light-cone Fock-state
wave functions.  The overlap of the two-particle
one-fermion--one-boson light-cone Fock state $|f \gamma>$
yields Schwinger's contribution~\cite{Schwinger} $a_e = {\alpha\over
2 \pi}$.  The light-cone
wave functions with $n \ge 2$ QED quanta contribute to $a_e$
beginning at order $\alpha^{n-1}$ as well as higher orders.  Thus a
systematic evaluation of the lepton's Fock-state wave functions as an
expansion in Fock number rather than perturbation theory would provide a
physically appealing answer to Feynman's challenge~\cite{hb}.

In principle the complete spectrum of a quantum field theory can be
determined by finding the eigenvalues of the light-cone
Hamiltonian~\cite{BPP}.
The Fock-state expansion of the eigensolutions at fixed light-cone time
$\tau = x^+=x^0 + x^z$  provides a frame-independent wave-function
description of the elementary and composite states in terms of the quanta
of the free Hamiltonian.
The discretized light-cone quantization (DLCQ) method
utilizes periodic boundary conditions
to truncate  the size of the Fock-state expansion while preserving boost
invariance.  This method has been successfully applied to a
large variety of gauge theories and supersymmetric theories in $1+1$ and
$2+1$ dimensions.

An essential problem in applying light-cone Fock-state methods to
renormalizable gauge theory is to regularize the calculations in such a
way as to preserve  Lorentz and gauge invariance, or at least preserve
them well enough to allow an effective renormalization to be performed.
Since the method of regularization must also allow for efficient
calculations to be performed, the problem presents a challenge.

We have recently performed nonperturbative calculations using the
generalized Pauli--Villars (PV) method  as an ultraviolet
regulator of (3+1)-dimensional quantum field theories.
We include a sufficient number of PV fields in the Lagrangian to
ensure that perturbation theory is finite.  This method explicitly
preserves Lorentz invariance;  in some cases, such as QED, it effectively
preserves gauge invariance.
In a case  where it breaks gauge invariance, such as QCD,
we have to add counterterms.
The PV regularization can produce a finite
theory which preserves Lorentz and gauge symmetries.  However,
if we do not have the exact solution, we must develop approximate
methods.  Our approximation involves truncating the
Fock space.  The truncation will break all of the symmetries.  However, the
usefulness of the truncated answer is a question of accuracy rather than
an issue of symmetry breaking.  With  regulators in place we presume that
there is an exact solution which preserves all symmetries including gauge
invariance, and if our approximate solution is close to the exact one,
even if the small difference is in such a direction as to maximally
violate the symmetries, it is still a small difference.  Of course, the
inclusion of negative-metric fields in the Lagrangian will also violate
unitarity.  We shall have more to say about this issue below.

The generalized PV method has been applied successfully
to  Yukawa-like theories~\cite{bhm1,bhm2,bhm3,bhm4,bhm5} where there are
no infrared divergences and no need to protect gauge symmetry.  An
important conclusion of these studies is that past
some threshold (which depends on the values of the coupling constant and
the values of the PV masses), there is always a rapid drop off
of the projection of the eigensolution wave function onto higher Fock
sectors, in contrast to the equal-time Fock-space expansion.
This provides a strong motivation for
the light-cone representation as a viable approximant to nonperturbative
theory.

In this paper we will test the  convergence  of the PV-regulated
light-cone Fock-state expansion for (3+1)-dimensional gauge theory
by applying it to
a nonperturbative calculation of the electron anomalous moment in QED.
In some ways this application to QED
is not an ideal test of these nonperturbative methods:
the physical electron is a very perturbative
object.  We do not expect to do better, or even as well as perturbation
theory.  However, that is not our objective; we simply want to
verify that an approximate nonperturbative solution for the electron's
magnetic moment is an approximation to QED.  Somewhat related work, from 
a strictly perturbative point of view, can be found in~\cite{lb}.

Careful studies have also
shown~\cite{Paston:1997hs,Paston:2000fq} that the
perturbative series obtained from particular combinations of PV
fields and higher derivative regulators
give  the same result as standard perturbation series regulated with
dimensional regularization.  These studies not only included Yukawa theory,
but also non-Abelian gauge theory.
In Section~\ref{sec:lc}, we shall apply this method of regularization,
together with a Fock-space truncation to the
calculation of the electron magnetic moment.  This is the first
application of  this regularization to a nonperturbative problem.

We find that there are three problems which must be solved in order
to produce a useful calculation of the electron's magnetic moment:
the problem of uncanceled divergences, the problem of maintaining
gauge invariance, and the problem of new singularities.  We believe
that we have found effective solutions to these problems, at least
for the present calculations.  The problem of uncanceled
divergences  occurs anytime we truncate the Fock space.
For example, if we truncate the
physical electron's Fock space to include only the subspace of one
fermion and one photon, calculate the
wave function nonperturbatively, and use that wave function to calculate
the moment, we obtain a result of the form:
\begin{equation} \label{eq:ae}
a_e = {\alpha \times {\rm [finite\; quantity]} \over 1
      + \alpha \times {\rm [finite\; quantity]} +
\alpha \times {\rm [finite\; quantity]} \log {\mu_1\over m^2_e}},
\end{equation}
where $\mu_1$ is the PV photon mass.
If we let $\mu_1$ become infinite,  we
will obtain a zero anomalous moment.

\begin{figure}[bhtp] \centerline{\includegraphics[width=9cm]{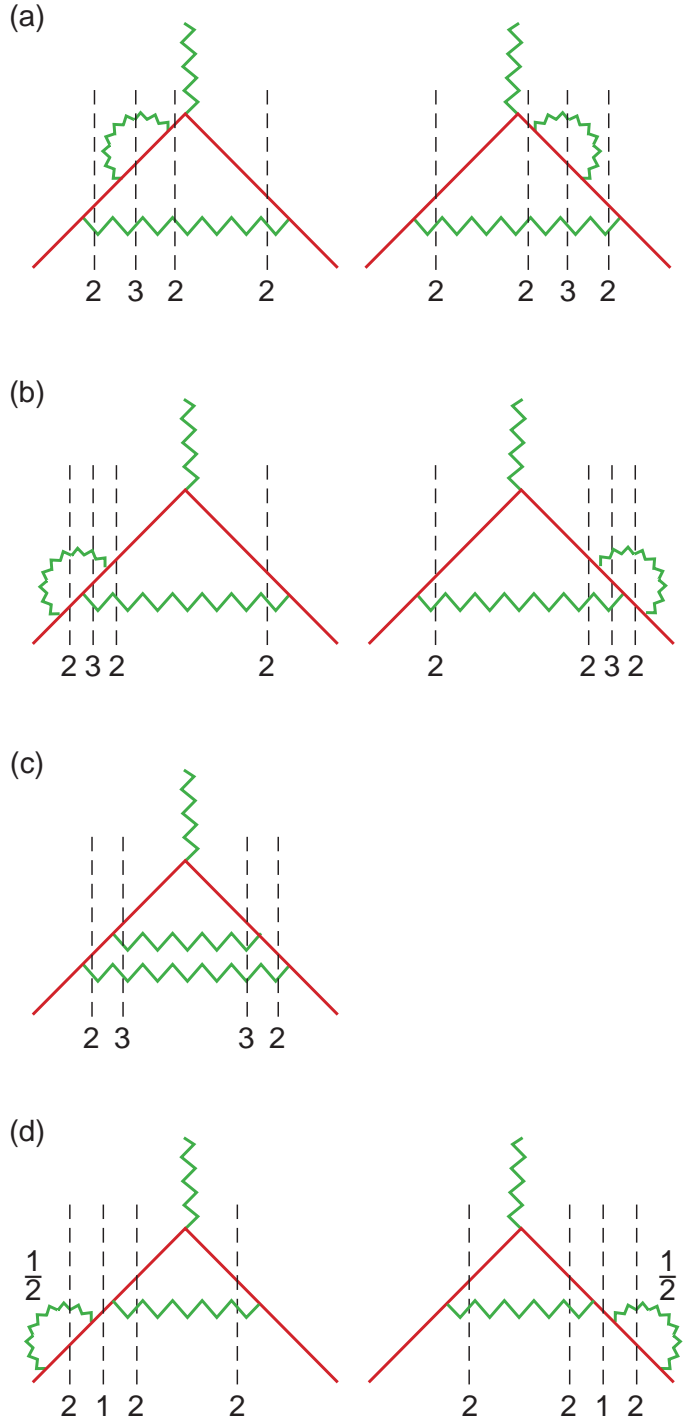}}
\caption{\label{fig:fymmX} Light-cone time-ordered contributions to the lepton
form factors, corresponding to the order-$\alpha^2$ ladder Feynman diagram in
perturbative QED. The vertical dashed lines mark intermediate states
with the indicated number of constituents.  For amplitudes associated
with diagrams in (d), with loops on the external legs, a factor of
1/2 is applied.}
\end{figure}

The origin of the  uncanceled divergence  in Eq.~(\ref{eq:ae})
can be seen by  examining the two-loop contributions to the electron
moment in perturbation theory.  The relevant double-ladder Feynman
diagrams are shown in Fig.~\ref{fig:fymmX}.
The Dirac $F_1(q^2)$ and Pauli $F_2(q^2)$ form factors
correspond to $S_z= \pm  S_z^\prime$ matrix elements, respectively.
(The $q^+=0$, $q^2_\perp=Q^2=-q^2$ frame is assumed.) Fock states with
particle number 1,2,3 contribute as indicated in the figure.  Only
time orderings with positive $k^+$ appear in light-cone quantization.
The Ward identity $Z_1=Z_2$ guarantees the cancellation of the divergences
from wave function and vertex renormalization subgraphs. However, if the
three-particle Fock state $|f \gamma \gamma> $ is excluded by the truncation
of the Fock space, $F_2$ is ultraviolet (UV) divergent since the vertex
correction shown in amplitude (c) is missing.  The divergent contribution
to the lepton anomalous moment $a_\ell = F_2(0)$ is the source of the UV
divergence in the denominator of Eq.~(\ref{eq:ae}) due to Fock-space
truncation. Of course, $F_1$ remains UV finite.
The divergence in Eq.~(\ref{eq:ae}) does not happen in perturbation theory;
since the numerator is already of order $\alpha$, we would use only the 1
from the denominator.  At order $\alpha^2$ there would be new terms in the
numerator which would cancel the divergent terms in the denominator of
Eq.~(\ref{eq:ae}).

The above discussion illustrates the problem of uncanceled
divergences.  While we could find ways of allowing the bare mass and the
coupling constant to depend on the PV masses to give
a finite expression, the results would not look
anything like the results from perturbation theory, since in QED
the coupling is renormalized only by vacuum polarization.  In addition, the
results would not make sense physically.  Our resolution of this
difficulty is to keep the PV masses finite.
The motivation is as follows: If the limit of
infinite PV masses would give a useful answer in the case
where we do not truncate the Fock space (so we have no uncanceled
divergences), then there must be some finite value of the PV
masses that would also give a useful answer.  The question is whether we
can use a sufficiently large value.  To answer that question we must
consider that there are two types of error associated with the values of
the PV masses.  The first type of error results in having
these masses too small; then our wave function will contain too much of
the negative-normed states, unitarity will be badly violated, and in the
worst case we might get negative probabilities.  We can roughly estimate
the magnitude of that type of error as
\begin{equation}
    E_1 = {M^2_0/M^2_1},
\end{equation}
where $M_0$ is the physical mass scale and $M_1$ is the
PV mass scale.  The other type of error results when
the PV masses are too large; in that case the true wave
function will project significantly onto the parts of the
representation space excluded by the truncation.
We can
roughly estimate the magnitude of that type of error as
\begin{equation}
    E_2 = {\langle \Phi_{+}^\prime|\Phi_{+}^\prime\rangle
\over \langle \Phi_{+}|\Phi_{+}\rangle},
\end{equation}
where $|\Phi_{+ }^\prime\rangle$ is the projection of the wave
function onto the excluded sectors.
In practice, the projection onto the first excluded Fock sector can
be estimated perturbatively using the projection of $P^-$ onto the higher
sectors as the perturbing operator.  Without additional information, the
best that we can do is to set $E_1$ equal to the perturbative estimate of
$E_2$.  Below, we will apply this procedure to the calculation of the
magnetic moment done in this paper. In general, if, at our estimated
optimum value for the PV mass scale both types of error are
small, we can do a useful calculation; otherwise, we cannot do a useful
calculation without expanding the part of the representation space that
we include in our calculation.

The main reason for believing that we can do a useful
calculation in spite of the problem of uncanceled divergences is
the lesson from the earlier studies mentioned above:
the observed rapid drop off
of the projection of the wave function onto higher Fock sectors.
Just where this rapid drop off occurs depends on the theory, the
coupling constant, and the values of the PV masses.  At
weak coupling and relatively light PV masses, only the
lowest Fock sectors are significantly populated.  At stronger
coupling or heavier PV masses, more Fock sectors will be
populated; but eventually the projection onto higher sectors will
fall rapidly.  The rapid drop off of the projection of the wave
function onto sufficiently high Fock sectors is the most important
reason why we do our calculations in the light-cone representation.
For any practical calculation in a realistic theory, we have to
truncate the space, and we must have a framework in which that
procedure can lead to a useful calculation.  The rapid drop off in
the projection of the wave function will not happen in the
equal-time representation, mostly due to the complexity of the vacuum
in that representation.

These features can be explicitly
demonstrated by setting the PV masses equal to the
physical masses.  In that case the theory becomes exactly
solvable~\cite{bhm4}.  The spectrum is the free spectrum, and the
theory is not useful for describing real physical processes due to
the strong presence of the negative-norm states in physical wave
functions; however, it still illustrates the points we have been trying
to make.  In the equal-mass case, the physical vacuum is the bare light-cone
vacuum, while it is a very complicated state in the equal-time
representation; the  physical wave functions project onto a finite
number of Fock sectors in the light-cone representation but onto an
infinite number of sectors in the equal-time representation.

As the PV masses become larger
than the physical masses, the light-cone wave functions project onto more of
the representation space.  This effect is increased as
the coupling constant becomes larger.  However, the wave
functions remain much simpler than in the equal-time representation,
and, to the extent we can do the calculations, there is always a
point of rapid drop off of the projection onto higher Fock sectors.
Due to this rapid drop off, we expect to find a PV
mass scale such that the error in the calculated value for a given
observable from the presence of negative-norm states and the
error from truncation can both be made arbitrarily small.
Therefore, we believe the
requirement to keep the value of the PV masses
finite does not impose a limit on the accuracy which
could, in principle, be achieved.
In practice, the size of the representation space may be too large for
presently available computing facilities.
Furthermore, for the method to be useful in practice,
there must not only be a value of the PV mass for which
both types of errors are small, but there must be a wide range of
such values since the optimum value for the PV mass can
only be rather crudely estimated.  A principal objective of the
present work is to test these ideas on a physically realistic
problem to which we know the answer.

The problem of maintaining gauge invariance turns out to be
nontrivial and quite instructive.  In the next section we consider
the following procedure: we take the light-cone evolution operator
$P^-$ for QED in light-cone gauge $A^+=0$
as constructed,  for example, in
Refs.~\cite{tbp}  and regulate it with PV
fields.  We looked at several cases:  PV photons alone,
PV fermions alone,  and a combination of both.  This procedure
does not lead to a viable nonperturbative formalism for the
electron magnetic moment;
furthermore, it does not lead to a correct calculation of the
electron self-energy, even at order $\alpha.$
The problem can be traced to a
failure to maintain gauge invariance.
The breakdown of the straightforward implementation of
the PV method in light-cone gauge shows that the successful
construction of the nonperturbative theory is nontrivial.

In Section~\ref{sec:fy} we perform the calculation in
Feynman gauge with one PV photon and one PV fermion;
this leads to a consistent formulation for the nonperturbative calculation
of the electron moment in QED.
In Section~\ref{sec:lc} we perform the calculation in light-cone gauge,
but use the more
sophisticated method of regulation proposed in~\cite{Paston:2000fq}.
We show that this method  provides a successful
formulation of the nonperturbative electron moment problem in light-cone
gauge with a result very similar to that in Feynman gauge.

In Sections~\ref{sec:fy} and \ref{sec:lc} we face the
problem of new singularities.  We have to do integrals with
denominators of the form $(- M^2 x (1-x) + m^2 x + \mu^2 (1-x) +
k_\perp^2)$, where $M$ is the physical electron mass, $m$ is the bare
electron mass, and $\mu$ is the photon mass.  When the bare mass is
less than the physical mass, as is the case in QED, there can be a
zero in this denominator.  In perturbation theory the expansion is
about $M = m$, and the denominator cannot vanish as long as the
photon is given a small nonzero mass.   The standard techniques in
perturbation theory thus avoid this singularity.  We find that when
the zero is a simple pole, the principal-value prescription is
correct.  However, in the wave function normalization the denominator is
squared, so there is a double pole, and we must give it a meaning.  We
propose the following prescription:
\begin{eqnarray}
   &&\int dy \; dk_\perp^2 \; 
   {f(y,k_\perp^2)\over [ m^2 y + \mu_0^2 (1-y) -M^2 y (1-y) + k_\perp^2]^2}
   \nonumber \\
&&\equiv
 \lim_{\epsilon\rightarrow 0}
{1\over 2 \epsilon} \int dy \int dk_\perp^2 f(y,k_\perp^2)
\Bigg[{1 \over [ m^2 y + \mu_0^2 (1-y) - M^2 y (1-y) + k_\perp^2 - \epsilon]}
   \nonumber \\
&&- {1 \over [ m^2 y + \mu_0^2 (1-y) - M^2 y (1-y) + k_\perp^2 +
\epsilon]}\Bigg],
\label{prescr}
\end{eqnarray}
where simple poles are prescribed as principal values.

This prescription has the interesting consequence that the wave
function normalization is infrared finite whereas it is infrared
divergent in perturbation theory.  Given this
prescription, the true singularity occurs at $M = m + \mu$; in
perturbation theory, with $M = m$, this is at $\mu = 0$, which is
the infrared singularity and the reason that the photon mass cannot
be taken all the way to zero in perturbation theory.  For the
nonperturbative calculation, the physical photon mass can be taken
to zero since $M \neq m$.  The basic requirement of these
prescriptions is that they preserve the Ward identities.
We anticipate using this prescription for
QCD where the basic requirement will be the preservation of the
Ward--Takahashi identities.  We have
not shown that the prescription preserves the Ward identities in QED,
but it does lead to a successful calculation in the present case.

A different approach to this same dressed-electron problem has
been taken by Karmanov, Mathiot, and Smirnov~\cite{Karmanov:2003hd}.
They use a covariant form of light-cone quantization without
Pauli--Villars regularization.  Their Hamiltonian then contains
instantaneous fermion interactions, and, in Feynman gauge, the
infinite number of terms generated by inversion of the covariant
derivative.  The problem of uncancelled divergences is avoided
by application of sector-dependent renormalization~\cite{Perry:mz}.
They must construct counterterms explicitly.
However, they truncate in a Fock basis where the constituent 
electron has the same mass as the dressed electron, bringing
their calculation closer to perturbation theory; in fact, they
find that any signifant difference with perturbation theory will
not appear until the basis is expanded to include higher Fock
states.  Also, they do not calculate the anomalous moment.

\section{Trouble in Light-Cone Gauge}
\label{sec:trouble}

In this and the next section
the notation that we use for light-cone coordinates is
\begin{equation}
x^\pm = x^0 \pm x^3\,,\;\; \vec{x}_\perp=(x^1,x^2).
\end{equation}
The time coordinate is $x^+$, and the dot product of two
four-vectors is
\begin{equation}
p\cdot x=\frac{1}{2}(p^+x^- + p^-x^+)
                -\vec{p}_\perp\cdot\vec{x}_\perp.
\end{equation}
The momentum component conjugate to $x^-$ is $p^+$, and the
light-cone energy is $p^-$.  Light-cone three-vectors are identified
by underscores, such as
\begin{equation}
\underline{p}=(p^+,\vec{p}_\perp).
\end{equation}
We use the following choice for the $\gamma$ matrices
\begin{eqnarray}
&\g^0=\ls
\begin{array}{cc}
I&0\\
0&-I
\end{array}
\rs,\qquad &\g^+=\ls
\begin{array}{cc}
I&\si_3\\
-\si_3&-I
\end{array}
\rs,\nonumber\\
&\g^-=\ls
\begin{array}{cc}
I&-\si_3\\
\si_3&-I
\end{array}
\rs,\qquad &\g^k=i\ls
\begin{array}{cc}
0&\si_k\\
-\si_k&0
\end{array}
\rs .
\end{eqnarray}
For additional details, see Appendix A of Ref.~\cite{bhm1}.

We use the standard $P^-$ for light-cone quantized QED in light-cone
gauge~\cite{tbp} with  modifications due to the inclusion of the
PV fields.  We should remark, however, that with
the inclusion of any number of PV Fermi fields, the four-point
interactions which would take a state of one electron and one
photon to another state of one electron and one photon are missing
from $P^-$; that such terms are not included below is not an
omission; the calculation is complete in our chosen subspace.  We
truncate the Fock space to the one-fermion sector plus the
one-fermion, one-photon sector.  We then solve the eigenvalue problem
\begin{equation}
    P^+ P^- \ket{s} = M^2 \ket{s},
\end{equation}
where the total $\vec{P}_\perp$ of the state is taken equal to 0.
As always with a Tamm-Dancoff truncation, we can solve for the wave
function in the highest Fock sector (one fermion plus one photon) by hand
and obtain an equation in the one-fermion sector.

We have regulated the theory in several ways.  We shall describe one
particular choice in some detail:  We use three PV Fermi fields with
flavor-changing currents.  The flavor-changing currents break gauge
invariance, and thus we might expect to require counterterms to correct
for that.  Nevertheless, we will proceed with the calculation without
counterterms.  That will allow us to study a case where a proper respect
for gauge invariance is not maintained.  Also, as we shall argue below,
the source of the breaking of gauge invariance is much deeper than that
due to the flavor-changing currents.

For the Lagrangian we take
\begin{equation}
  -{1 \over 4}F^{\mu \nu} F_{\mu \nu}
  + \sum_{i=0}^3 {1 \over \nu_i}\bar{\psi_i}^\prime (i \gamma^\mu \partial_\mu
  - m_i) \psi_i^\prime - e \bar{\psi}^\prime\gamma^\mu \psi^\prime A_\mu,
\end{equation}
where
\begin{equation}
   \psi^\prime = \sum_{i=0}^3 \psi_i^\prime , \quad \sum_{i=0}^3 \nu_i  = 0 ,
    \quad \sum_{i=0}^3 \nu_i  m_i = 0 ,   \quad \sum_{i=0}^3 \nu_i  m_i^2 = 0.
\end{equation}
A particular realization of these PV conditions is
\begin{equation}
  \{\nu_i\} = \{1;3;-1;-3\} , \quad \{m_i\} =
       \{m_0;-m_0 + 2 m_3;-2m_0 + 3 m_3;m_3\}.
\end{equation}
With this choice it is convenient to define
\begin{equation}
\psi_0 = \psi_0^\prime , \quad \psi_1 = \sqrt{3} \psi_1^\prime, \quad
\psi_2 = \psi_2^\prime , \quad \psi_3 = \sqrt{3} \psi_3^\prime ,
\end{equation}
so that the $\psi$ fields are canonically normalized (except for the
minus signs for $\psi_2$ and $\psi_3$).  The coupling to the $A$ field is
\begin{equation}
  e \bar{\psi}^\prime\gamma^\mu \psi^\prime A_\mu
      = \sum_{i j} g_{i j}\bar{\psi}_i\gamma^\mu \psi_j A_\mu,
\end{equation}
where
\begin{eqnarray}
g_{0 0}&=&  g_{0 2}= g_{2 2}= e , \quad
g_{0 1}= g_{0 3}= g_{1 2}= g_{23}= \sqrt{3} e , \\
g_{1 1}&=&  g_{1 3}= g_{3 3}= 3 e , \nonumber
\end{eqnarray}
and $g_{i j} = g_{j i}$.

We use the mode expansions
\begin{equation}
\psi_{i +}(\ub{x})={1\over\sqrt{16\pi^3}} \sum_{s} \int d\ub{k} \;
\chi_s \; \Bigl[ b_{s}(i,\ub{k}) e^{-i\ub{k}\cdot\ub{x}}+
d^\dagger_{-s}(i,\ub{k})e^{+i\ub{k}\cdot\ub{x}}\Bigr], \label{me1}
\end{equation}
\begin{equation}
A^i(\ub{x})={1\over\sqrt{16\pi^3}} \sum_{\lambda} \int d\ub{k} \;
 \; {1 \over \sqrt{k^+}}\Bigl[ a_{\lambda}(\ub{k})
\epsilon^i(\lambda) e^{-i\ub{k}\cdot\ub{x}}+ a_{\lambda}^\dagger(\ub{k})
{\epsilon^i(\lambda)}^*e^{+i\ub{k}\cdot\ub{x}}\Bigr]\label{me2},
\end{equation}
where the polarization states are
\begin{equation}
\chi_{+\ha}={1\over\sqrt2}\left(\matrix{ 1\cr0\cr1\cr0\cr}\right),
\qquad\qquad \chi_{-\ha}={1\over\sqrt2}\left(\matrix{
0\cr1\cr0\cr-1\cr}\right),
\end{equation}
\begin{equation}
{\e}_{\perp,+1}\equiv{-1\over\sqrt{2}}(1,i),  \qquad\qquad
  {\e}_{\perp,-1}\equiv {1\over\sqrt{2}}(1,-i).
\end{equation}
The vertex functions can be computed from $P^-$ as
\begin{equation}
   V_{i j} = {\epsilon_{i j} \over \sqrt{16 \pi^3}}
       {1 \over \sqrt{1-x}}{1 \over x(1-x)}(k_x - i k_y),
\end{equation}
\begin{equation}
   W_{i j} = {\epsilon_{i j} \over \sqrt{16 \pi^3}}
       {1 \over \sqrt{1-x}}{1 \over (1-x)}(-k_x - i k_y),
\end{equation}
\begin{equation}
   U_{i j} = {\epsilon_{i j} \over \sqrt{16 \pi^3}}
       {1 \over \sqrt{1-x}} (-m_i + {m_j \over x}),
\end{equation}
\begin{equation}
   \tilde{U}_{j i} = {\epsilon_{j i} \over \sqrt{16 \pi^3}}
        {1 \over \sqrt{1-x}} (-m_i + {m_j \over x}),
\end{equation}
where
\begin{equation}
   \epsilon_{i j} \equiv \cases{g_{i j},&if $i = 0,1$;\cr
        -g_{i j},&if $i = 2,3$.\cr}
\end{equation}

We expand the eigenstate as
\begin{eqnarray}
 |s> = &&\sum_{i=0}^3 z_i b_+^\dagger(i,\ub{P}) |0>
    + \sum_{i=0}^3   \int d\ub{k} f(i,x,\vec{k}_\perp) b_+^\dagger(i,\ub{k})
                      a_+^\dagger(\ub{P}-\ub{k}) |0> \nonumber \\
+ &&\sum_{i=0}^3 \sum_{s=\{-,+\}} \int d\ub{k} g_s(i,x,\vec{k}_\perp)
b_s^\dagger(i,\ub{k}) a_{-s}^\dagger (\ub{P}-\ub{k}) |0> .
\end{eqnarray}
We will take the wave function normalization $z_0$ to be 1 for the
moment and calculate it later.
{}From $P^+ P^- |s> = m_e^2 |s>$ we find immediately (in units where
$m_e$ is taken to 1):
\begin{equation}
   f(i,x,\vec{k}_\perp)
     = { \sum_{j=0}^3  z_j V_{j i}(x,\vec{k}_\perp) \over 1
      - {m_i^2 + k_\perp^2 \over x} - {\mu^2 + k_\perp^2 \over 1-x}} ,
\end{equation}
\begin{equation}
   g_+(i,x,\vec{k}_\perp)
     = { \sum_{j=0}^3  z_j W_{j i}(x,\vec{k}_\perp) \over 1
      - {m_i^2 + k_\perp^2 \over x} - {\mu^2 + k_\perp^2 \over 1-x}} ,
\end{equation}
\begin{equation}
   g_-(i,x,\vec{k}_\perp)
     = { \sum_{j=0}^3  z_j U_{j i}(x,\vec{k}_\perp) \over 1
      - {m_i^2 + k_\perp^2 \over x} - {\mu^2 + k_\perp^2 \over 1-x}} .
\end{equation}
{}From all this we derive the four nonlinear equations
\begin{eqnarray}
  z_i m_i^2 + &&\sum_{j=0}^3 \int dxd^2k_\perp {V_{j i}^*(x,\vec{k}_\perp)
      \sum_{l=0}^3 z_l V_{l j}(x,\vec{k}_\perp) \over 1
         - {m_j^2 + k_\perp^2 \over x} - {\mu^2 + k_\perp^2 \over 1-x}}
         \nonumber \\
+ &&\sum_{j=0}^3 \int dxd^2k_\perp {W_{j i}^*(x,\vec{k}_\perp)
     \sum_{l=0}^3 z_l W_{l j}(x,\vec{k}_\perp) \over 1
         - {m_j^2 + k_\perp^2 \over x} - {\mu^2 + k_\perp^2 \over 1-x}}
         \nonumber  \\
+ &&\sum_{j=0}^3 \int dxd^2k_\perp {\tilde{U}_{j i}^*(x,k_\perp)
\sum_{l=0}^3 z_l U_{l j}(x,k_\perp) \over 1 - {m_j^2 + k_\perp^2
\over x} - {\mu^2 + k_\perp^2 \over 1-x}} = z_i .
\end{eqnarray}
These can be written in terms of the three integrals:
\begin{equation}
   J =  \int dx dz \sum_{i=0}^3 {1 \over x} {({1 + x^2 \over (1 - x)^2} z
       + m_i^2) \nu_i \over x (1-x) - m_i^2 (1-x) - \mu^2 x - z} ,
\end{equation}
\begin{equation}
   I_1 = \int dx dz \sum_{i=0}^3  {m_i \nu_i \over x (1-x) - m_i^2 (1-x)
       - \mu^2 x - z} ,
\end{equation}
\begin{equation}
   I_0 = \int dx dz \sum_{i=0}^3  {x \nu_i \over x (1-x) - m_i^2 (1-x)
        - \mu^2 x - z} ,
\end{equation}
where $z = k^2_\perp.$
The nonlinear equations become
\begin{eqnarray}
\lefteqn{z_i m_i^2 + {e^2 \over 16 \pi^2}J \sum_{j=0}^3 \epsilon_{j i} z_j
    - {e^2 \over 16 \pi^2}I_1 \sum_{j=0}^3 \epsilon_{j i} z_j (m_i + m_j) }&&
        \nonumber \\
&&+{e^2 \over 16 \pi^2}I_0 \sum_{j=0}^3 \epsilon_{j i} z_j m_i m_j = z_i .
\end{eqnarray}
While these equations are complicated, they can be simplified by the
observation that, for the parameter values of interest, $J$
is very much larger than $I_0$ and $I_1$.  Also, for those
parameter values, $z_1$, $z_2$, and $z_3$ are small compared
to one, and a solution sufficiently accurate for our needs is given
by the simple expression
\begin{equation}
   m_0^2 = 1 - \frac{\alpha}{4\pi} J.
\end{equation}

We can use the wave function to calculate the anomalous
magnetic moment using the formalism of Brodsky and
Drell~\cite{Brodsky:1980zm}.  The contribution from the physical
field is much larger that the contributions from the PV
fields, so we have
\begin{equation}
   a_e = {\alpha \over \pi} z_0^2 \int dx {m_1 x^2 (1-x) \over m_1^2 x
   + \mu^2 (1-x) - x (1-x)} ,
\end{equation}
where $z_0$ is determined by wave function normalization:
\begin{eqnarray}
   1/z_0^2 &=& 1 + z_1^2 - z_2^2 - z_3^2 \\
+ &\int& dxd^2k_\perp
 (|f(0,x,\vec{k}_\perp)|^2 + |f(1,x,\vec{k}_\perp)|^2
        - |f(2,x,\vec{k}_\perp)|^2  - |f(3,x,\vec{k}_\perp)|^2)\nonumber \\
+ &&(|g_+(0,x,\vec{k}_\perp)|^2 + |g_+(1,x,\vec{k}_\perp)|^2
   - |g_+(2,x,\vec{k}_\perp)|^2 - |g_+(3,x,\vec{k}_\perp)|^2)\nonumber \\
+ &&(|g_-(0,x,\vec{k}_\perp)|^2 + |g_-(1,x,\vec{k}_\perp)|^2
   - |g_-(2,x,\vec{k}_\perp)|^2 - |g_-(3,x,\vec{k}_\perp)|^2) .\nonumber
\end{eqnarray}
We find
\begin{equation}
     a_e = {\alpha \over 2 \pi} z_0^2
       ({1 \over 2}
       +  \frac{\alpha}{4\pi} J(1 + \frac{\alpha}{4\pi} J)
                 \ln({4\pi \over \alpha J} - 1) - \frac{\alpha}{4\pi} J) .
\end{equation}

We can now state the problem with this calculation: $J$ has a very
strong dependence on $m_3$ (the PV mass scale).  That,
in turn, gives our estimate of the anomalous magnetic moment a very
strong dependence on $m_3$.  If we use units of ${\alpha \over 2
\pi}$, so that the correct value is near one, then we find that, even
with a value for the photon mass as large as 0.5 electron masses,
when $m_3$ changes from 3 times the electron mass to 7 times the
electron mass,  $a_e$ changes from 1.2 to -1.2.  If we use a
smaller value for the photon mass, which we would surely have to do
to get useful results, the dependence is even stronger.  Since we
cannot hope to estimate the optimum value for the PV
mass scale even to within this range, the present calculation is
clearly useless.  The problem is clearly the loss of
gauge invariance; gauge invariance should prevent such strong
behavior.  One might reasonably think that the problem is the 
flavor-changing currents we have included in the calculation, but we 
shall now argue that the worst breaking of gauge invariance has a more
fundamental source.  We shall return to the breaking due to 
flavor-changing currents in the next section.

We note that if we keep only the physical field and set $M = m_0
=m$, the function which appears in our nonlinear equations is just
the (unregulated) one-loop fermion self-energy
\begin{equation}
    \frac{\alpha}{4\pi}\left( J - 2 I_1 + I_0 \right)
    = \frac{\alpha}{4\pi} \int dx dz  {1 \over x}
    {{1 + x^2 \over (1 - x)^2} z + m^2 (1-x)^2  \over m^2 x (1-x)
                  - m^2 (1-x) - \mu^2 x - z} \label{lca} .
\end{equation}
Therefore, a very useful point of comparison is the evaluation
of the fermion self-energy  calculated in the paper by
Brodsky, Roskies and Suaya~\cite{brs}, hereafter referenced as BRS.
They evaluated all the graphs needed to calculate the electron's
magnetic moment in  perturbation theory
through order $\alpha^2$.
Included in their calculations is the one-loop electron self-energy.
They did not use light-cone quantization but wrote down time-ordered
perturbation theory in the equal-time representation and then boosted to
the infinite momentum frame.  They worked in Feynman gauge, but the
electron self-energy should be gauge invariant.  They give the
self-energy as the sum of two separate pieces, $m_a$ and $m_b$;
$m_b$ results from boosting the z-graph while $m_a$ comes from the
other time ordering.  They obtain in our notation (see Eqs.~(3.40)
and (3.41) of BRS)
\begin{equation}
    m_a = {e^2 \over 16 \pi^2} \int dx dz {1\over x}
       {z + m^2(1 - 4 x - x^2) \over m^2 x (1-x) - m^2 (1-x) - \mu^2 x - z} ,
\end{equation}
\begin{equation}
    m_b = - {e^2 \over 16 \pi^2}\int dz \ln(\mu^2 + z) .
\end{equation}
BRS show that if the theory is regulated with the inclusion of two
PV photons, the sum of $m_a$ and $m_b$ is equal to the
usual Feynman one-loop self-energy in QED, regulated in the same
way.  Notice that $m_b$ implements the requirement of chiral
symmetry:  the shift in the bare mass is zero if the bare mass is
zero, that is, $ \delta m_b = -\delta m_a \bigm|_{m=0}$.  Terms
such as $m_b$ are often missed in light-cone quantization.  That fact
has been noticed at least as far back as the work of Chang and Yan~\cite{cy}.
Those authors suggest that the problem can be solved by including an
extra PV field and using it to implement the chiral
symmetry condition; that is a technique we have used in the
past~\cite{bhm1}.  Since $m_b$ does not depend on the mass,
inclusion of PV fermi fields will also solve the problem.
That possibility was noticed by BRS, and we shall use this method in
the next sections.  Therefore without PV fermi fields
we might not expect to get the
sum of $m_a$ and $m_b$ correctly, but we should expect to at least
reproduce $m_a$.  We
do not have to get the same integrand, but we should get the same
result after integration if the regulation preserves covariance.
However, if we use two PV photons, or three PV
photons with the third field used to eliminate $m_b$,
we do not obtain the correct result for $m_a.$

We can gain some further insight into what is going on by observing
that, in Feynman methods, the light-cone gauge is obtained from
Feynman gauge by the replacement
\begin{equation}
   g^{\mu\nu} \rightarrow g^{\mu\nu}
       - {n_\mu k_\nu + n_\nu k_\mu \over n\cdot k} .
\end{equation}
{}From that replacement there is no obvious source for the double pole
at $x=1$ that we see in our light-cone calculation.  The explanation
is that, in some formal sense without worrying about regulation,
light-cone quantization accomplishes the replacement by writing
\begin{equation}
g^{\mu\nu} \rightarrow g^{\mu\nu} - {n_\mu k_\nu + n_\nu k_\mu \over
n\cdot k} + {k^2 n^\mu n^\nu\over (n \cdot k)^2} - {k^2 n^\mu
n^\nu\over (n \cdot k)^2} .
\end{equation}
The first three terms correspond to the usual three-point
interactions that one obtains by usual light-cone quantization as given, for
instance, in~\cite{tbp}.  The last term is given by the so called
``instantaneous'' (four-point) interactions which result from
solving the constraint equation for the photon.  In at least some
cases, for tree level processes the required cancellations actually
occur, and results equivalent to the equal-time formulation are
obtained.  However, here, at one loop, the cancellations are not working
correctly, and we can see why as follows:  If we write down the
Feynman integral whose numerator is
\begin{equation}
 - {k^2 n^\mu n^\nu\over (n \cdot k)^2} ,
\end{equation}
regulate the calculation with two PV photons, and perform
the $k^-$ integral, we get an amplitude which cannot be obtained
from a four-point interaction
\begin{equation}
     2 \int dx dz {1\over x}{z({1\over 1-x} -{1\over (1-x)^2})  -
    m^2 x \over m^2 x (1-x) - m^2 (1-x) - \mu^2 x - z} .
\end{equation}
When this amplitude is added to our light-cone amplitude,
(\ref{lca}), we get precisely $m_a$.  There is no doubt that $m_a$
--- with the perturbative denominator, $m^2 x (1-x) - m^2 (1-x) -
\mu^2 x - z$, replaced by the nonperturbative denominator, $M^2 x
(1-x) - m^2 (1-x) - \mu^2 x - z$ --- is the function that should
enter our nonlinear equations.  We could fix the problem here by
just using $m_a$ with the perturbative denominator replaced
by the nonperturbative denominator, but since we do not know how to
make similar corrections in other cases, we do not consider that
replacement to be useful.  It is clear that the problem is that gauge
invariance has been lost in solving the constraint equation
\begin{equation} \label{eq:constraint}
 \partial_-^2 A^- + \partial_-\partial_i A^i = -e \Psi_+^\dagger \Psi_+ .
\end{equation}
It is possible that the wrong boundary conditions
have been used in solving this equation
or that the equation must be modified:
the constraint equation satisfied by the regulated fields and something like
Schwinger terms may need to be included. The loss of gauge invariance
using standard light-cone techniques deserves further study.
We will now turn our attention to the use of other
gauges and other methods of regulation.

In the next section we will discuss light-cone-gauge quantization using 
Feynman gauge and PV regularization. In Sec.~\ref{sec:lc} we shall show that 
a successful calculation can also be made in light-cone gauge, if the 
formalism is augmented with higher derivative regulators and several 
Pauli--Villars fields.  In that case, due to the
higher derivatives, $A^-$ is a degree of freedom, and there is no
equivalent of Eq.~(\ref{eq:constraint}).

\section{Feynman Gauge}
\label{sec:fy}
In this section we shall calculate the electron's
magnetic moment using light-cone quantization in Feynman gauge.
We shall regulate the theory
by the use of one PV photon and one PV
fermion with the inclusion of flavor-changing currents.  The
Lagrangian is thus
\begin{eqnarray}\label{eq:fgLagrangian}
  &&\sum_{i=0}^1 \left(-{1 \over 4} (-1)^i F_i^{\mu \nu} F_{i,\mu \nu} +
 (-1)^i \bar{\psi_i} (i \gamma^\mu \partial_\mu - m_i) \psi_i
 +  B_i \partial_\mu A_{i}^{\mu} + {1 \over 2} B_i B_i \right) \nonumber \\
&&- e \bar{\psi}\gamma^\mu \psi A_\mu ,
\end{eqnarray}
where
\begin{equation}
  A^\mu  = \sum_{i=0}^1 A^\mu_i ,  \quad \psi =
\sum_{i=0}^1 \psi_i , \quad F_i^{\mu \nu}
    = \partial^\mu A_{i}^{\nu}-\partial_\nu A_{i}^{\mu}.
\end{equation}
Here, $i=0$ indicates the physical fields, and $i=1$, the PV
(negative-metric) fields.

We will now discuss two important consequences of including PV Fermi
fields with flavor-changing currents, one good effect and one
apparently bad effect.  The good effect pertains to
the operator $P^-$.  If one works out $P^-$ including only the
physical fields, one encounters the need to invert the covariant
derivative~\cite{sb} $\partial_- - e A_-$.  The same problem occurs
in any gauge where $A_-$ is not zero.  This complication is perhaps
the main reason that gauges other than light-cone gauge have
received relatively little attention in the light-cone
representation.  While the inverse of the covariant derivative can
be approximately defined by a power series in $e$, or, in a truncated
space may be calculated exactly if the truncation is sufficiently severe,
it is not clear that $P^-$ has been fully specified.  However, with
the inclusion of the PV fermions with flavor-changing currents, this
problem does not occur: the inverse of the covariant derivative is replaced
by the inverse of the ordinary derivative.  The part of $P^-$ that we shall
need in our calculations is given by
\begin{eqnarray} \label{eq:QEDP-}
\lefteqn{P^-=
   \sum_{i,s}\int d\ub{p}
      \frac{m_i^2+p_\perp^2}{p^+}(-1)^i
          b_{i,s}^\dagger(\ub{p}) b_{i,s}(\ub{p})} \\
   && +\sum_{l,\mu}\int d\ub{k}
          \frac{\mu_l^2+k_\perp^2}{k^+}(-1)^l\epsilon^\mu
             a_l^{\mu\dagger}(\ub{k}) a_l^\mu(\ub{k})
          \nonumber \\
   && +\sum_{i,j,l,s,\mu}\int d\ub{p} d\ub{q}\left\{
      b_{i,s}^\dagger(\ub{p}) \left[ b_{j,s}(\ub{q})
       V^\mu_{ij,2s}(\ub{p},\ub{q})\right.\right.\nonumber \\
      &&\left.\left.\rule{0.5in}{0in}
+ b_{j,-s}(\ub{q})
      U^\mu_{ij,-2s}(\ub{p},\ub{q})\right] a_{l\mu}^\dagger(\ub{q}-\ub{p})
                    + h.c.\right\}\,,  \nonumber
\end{eqnarray}
where $\epsilon^\mu = (-1,1,1,1)$ and
\begin{eqnarray}
    V^0_{ij\pm}(\ub{p},\ub{q}) &=& {e \over \sqrt{16 \pi^3 }}
                   \frac{ \vec{p}_\perp\cdot\vec{q}_\perp
                      \pm i\vec{p}_\perp\times\vec{q}_\perp
                       + m_i m_j + p^+q^+}{p^+q^+\sqrt{q^+-p^+}} , \\
    V^3_{ij\pm}(\ub{p},\ub{q}) &=& {-e \over \sqrt{16 \pi^3}}
                        \frac{ \vec{p}_\perp\cdot\vec{q}_\perp
                      \pm i\vec{p}_\perp\times\vec{q}_\perp
                       + m_i m_j - p^+q^+ }{p^+q^+\sqrt{q^+-p^+}} , \\
    V^1_{ij\pm}(\ub{p},\ub{q}) &=& {e \over \sqrt{16 \pi^3}}
       \frac{ p^+(q^1\pm i q^2)+q^+(p^1\mp ip^2)}{p^+q^+\sqrt{q^+-p^+}} , \\
    V^2_{ij\pm}(\ub{p},\ub{q}) &=& {e \over \sqrt{16 \pi^3}}
       \frac{ p^+(q^2\mp i q^1)+q^+(p^2\pm ip^1)}{p^+q^+\sqrt{q^+-p^+}} , \\
    U^0_{ij\pm}(\ub{p},\ub{q}) &=& {\mp e \over \sqrt{16 \pi^3}}
       \frac{m_j(p^1\pm ip^2)-m_i(q^1\pm iq^2)}{p^+q^+\sqrt{q^+-p^+}} , \\
    U^3_{ij\pm}(\ub{p},\ub{q}) &=& {\pm e \over \sqrt{16 \pi^3}}
       \frac{m_j(p^1\pm ip^2)-m_i(q^1\pm iq^2)}{p^+q^+\sqrt{q^+-p^+}} , \\
    U^1_{ij\pm}(\ub{p},\ub{q}) &=& {\pm e \over \sqrt{16 \pi^3}}
                            \frac{m_iq^+-m_jp^+ }{p^+q^+\sqrt{q^+-p^+}} , \\
    U^2_{ij\pm}(\ub{p},\ub{q}) &=& {i e \over \sqrt{16 \pi^3}}
                     \frac{m_iq^+-m_jp^+ }{p^+q^+\sqrt{q^+-p^+}} .
\end{eqnarray}

The apparently bad effect of the flavor-changing currents is that they break
gauge invariance.  That would seem to require the inclusion of
counterterms in the Lagrangian to correct for the symmetry breaking.
It turns out that  such counterterms are  not necessary:
as we shall see, we can take the limit
of the PV fermion mass $m_1 \to \infty$.  One might properly
worry that there might still be residual, finite effects of the necessary
counterterms, but the counterterms go to zero as powers of $m_1$ while
the only divergences we encounter are logs.  We shall therefore
proceed with the calculation using only the Lagrangian given above.

We use mode expansions similar to those of (\ref{me1}) and
(\ref{me2}) and expand the wave function as
\begin{equation}
  |\psi\rangle = \sum_i z_i b_{i,+}(\ub{P}) |0\rangle + \sum_{s,\mu,i,l}
 \int d\ub{k}C^\mu_{s,i,l}(\ub{k}) b_{is}^\dagger(\ub{k})
               a^\dagger_{l\mu}(\ub{P}-\ub{k}) |0\rangle .
\end{equation}
We set the total transverse momentum of the state to zero.
We can solve for the $C$'s as
\begin{eqnarray}
C^\mu_{+,i,l}(\ub{k}) &=& {\sum_j (-1)^j z_j P^+ V_{ij+}^\mu(\ub{k},\ub{P})
    \over
     M^2 - {m_i^2 + k_\perp^2 \over x} - {\mu_l^2 + k_\perp^2 \over 1-x}} , \\
C^\mu_{-,i,l}(\ub{k}) &=& {\sum_j (-1)^j z_j P^+ U_{ij+}^\mu(\ub{k},\ub{P})
     \over
     M^2 - {m_i^2 + k_\perp^2 \over x} - {\mu_l^2 + k_\perp^2 \over 1-x}} .
\end{eqnarray}

The eigenvalue equations for $z_i$ become
\begin{eqnarray}
(M^2-m_i^2)z_i &=& \int dx\; d^2k_\perp \sum_{\mu,i',j,l}(-1)^{i'+j+l}
z_j (P^+)^3 \epsilon^\mu \\
 && \times  {V^\mu_{i'j+}(\ub{k},\ub{P}) V^{\mu*}_{i'i+}(\ub{k},\ub{P}) 
             +U^\mu_{i'j+}(\ub{k},\ub{P})U^{\mu*}_{i'i+}(\ub{k},\ub{P})  \over
      M^2 - {m_{i'}^2 + k_\perp^2 \over x} - {\mu_l^2 + k_\perp^2 \over 1-x}} ,
      \nonumber
\end{eqnarray}
which can be written more usefully as
\begin{eqnarray} \label{eq:FeynEigen}
(M^2-m_i^2)z_i &=&
      2e^2\sum_j (-1)^j\left[M^2 z_j \bar{J}+m_iz_jm_j \bar{I}_0 \right. \\
 &&\left. \rule{1in}{0mm} -2Mz_j(m_i+m_j) \bar{I}_1 \right], \nonumber
\end{eqnarray}
with
\begin{eqnarray}
\bar{I}_n&=&\int\frac{dx dk_\perp^2}{16\pi^2}
   \sum_{jl}\frac{(-1)^{j+l}}{M^2-\frac{m_j^2+k_\perp^2}{x}
                                   -\frac{\mu_l^2+k_\perp^2}{1-x}}
   \frac{(m_j/M)^n}{(1-x)x^n}\,, \\
\bar{J}&=&\int\frac{dx dk_\perp^2}{16\pi^2}
   \sum_{jl}\frac{(-1)^{j+l}}{M^2-\frac{m_j^2+k_\perp^2}{x}
                                   -\frac{\mu_l^2+k_\perp^2}{1-x}}
   \frac{(m_j^2+k_\perp^2)/M^2}{(1-x)x^2} .
\end{eqnarray}
This form matches that of the equivalent eigenvalue problem in Yukawa
theory~\cite{bhm5}, with $M$ used as the mass scale instead of
$\mu_0$ and with the replacements $g^2\rightarrow 2e^2$ and
$I_1\rightarrow -2\bar{I}_1$.  The integrals $\bar{I}_0$ 
and $\bar{J}$ are equal.  
The solution to the eigenvalue problem can be transcribed from
 \cite{bhm5}; it is
\begin{equation}
e^2=\frac{(M\mp m_0)(M\mp m_1)}{2M(m_1-m_0)(2 \bar{I}_1\mp \bar{I}_0)} , \;\;
z_1=\frac{M \mp m_0}{M \mp m_1}z_0 ,
\end{equation}
with $z_0$ determined by normalization.

We now must discuss the problem to which we alluded earlier: the appearance
of new singularities.  Since $m_0 < 1$, and for the parameter values
of interest $m_0 > -1$, there will be a pole in the integrand
for the $i = l = 0$ term in the equation.  The effect looks like a
threshold, but there  is  no state into
which the system can decay.  The fact that the pole can exist is due
to the indefinite metric.  We believe that it is an artifact, and
that the correct procedure is to define the singularity
using the principal-value prescription.
That is what we shall do for the eigenvalue equation, but the same
singularity occurs in the normalization integral and in the equation
for the magnetic moment.  In those cases it is a double pole, and we
cannot define it as a principal value.  We shall return to this
point presently.  Interpreting the singularity in (\ref{eq:FeynEigen})
as a principal value, we can perform the integrations and then take
the limits of the PV mass $m_1\rightarrow \infty$ and
the physical photon mass $\mu_0 \rightarrow 0$. 
The complete result is quite complicated, but for the parameter values
of interest we can find several much simpler
approximate expressions for $m_0$.  One is given by
\begin{equation}
      m_0^2 \approx 1 - 6\, \frac{\alpha}{4\pi} \frac{1 + 2 \ln[\mu_1^2]}
             {2 - \frac{5\alpha}{4\pi} + \frac{7\alpha}{2\pi} \ln[\mu_1^2]} .
\end{equation}
Here we set the physical electron mass $M$ to 1.
For all except the largest values of $\mu_1$ in the range of
interest, we can just use
\begin{equation}
    m_0^2 \approx 1 - \frac{3\alpha}{4\pi} (1 + 2 \ln[\mu_1^2])
    \label{m1} \label{eq:m0f}.
\end{equation}

We must now calculate the integral that appears in the normalization condition;
it is given by
\begin{equation}
\frac{\alpha}{2\pi}  \int dx\; dk_\perp^2 \sum_{i,j}(-1)^{i + j} (1- x)
\left[{(m_j^2 - 4 m_0
m_j x + m_0^2 x^2) + k_\perp^2 \over  (x (1-x) - m_j^2 (1-x) - \mu_i^2 x  -
k_\perp^2)^2}\right] .
\end{equation}
The $i = j =0$ term of this integral contains the double pole mentioned
earlier.  Using the prescription discussed in the
Introduction, we will define the normalization integral as
\begin{eqnarray}
     && \frac{\alpha}{2\pi} \lim_{\epsilon\rightarrow 0}
                         {1\over 2 \epsilon}  \int dx\; dz \left\{
(1-x) \left(m_0^2(1 - 4  x + x^2) + z\right)  \right. \nonumber \\
&& \times\Bigl[{1 \over [ m_0^2 (1-x) + \mu_0^2 x - x (1-x) + z - \epsilon]}
\nonumber \\
&&-{1 \over [ m_0^2 (1-x) + \mu_1^2 x - x (1-x) + z + \epsilon]}\Bigr]
\nonumber \\
&&\left.
+\sum_{i \& j \neq 0}(-1)^{i + j} (1- x)\left[{(m_j^2 - 4 m_0 m_j x
+ m_0^2 x^2) + z \over  (x (1-x) - m_j^2 (1-x) - \mu_i^2 x  -
z)^2}\right] \right\},
\end{eqnarray}
with $z=k_\perp^2$.

This is the closest we can come to defining the double pole as the
derivative of a simple pole.  The major requirement of the
prescription for the double pole is that, combined with the
principal-value prescription for the single pole, it should respect
the Ward identity.     When,  we calculate the magnetic moment below,
we shall also
find a double pole in an integral and shall specify a meaning for it
in a similar way.  The
double-pole prescription does affect the  calculation of the magnetic moment;
nevertheless, the prescription allows the
calculation to proceed along lines closely parallel to those of
perturbation theory, and the value of the anomalous moment
is consistent with perturbation theory.  We therefore believe that the
prescriptions we have given are consistent,  but
we have not explicitly verified that they respect the Ward identity.

An interesting consequence of the double-pole prescription is
that the
normalization integral is now infrared finite: after performing the
integrations we can take the limit $\mu_0 \rightarrow 0$.  We did
not expect this, since in perturbation theory the wave function
renormalization constant is infrared divergent.  The difference can
be traced to the fact that, with the double-pole prescription, the true
singularity in the normalization integral is at $M = m + \mu$.
Since perturbation theory is an expansion around $M=m$, the
singularity in perturbation theory is always at $\mu = 0$.  In the
nonperturbative
calculation, the integral involving the physical fields has $\mu_0
= 0 \neq M - m_0$, while the integrals involving the
PV fields  do not encounter the double pole at all.
The ability to take the mass of the physical photon to zero is a
useful advantage in performing an approximate calculation of the
electron's magnetic moment because the anomalous moment is quite
sensitive to a nonzero photon mass~\cite{hb}.

\begin{figure}[bhtp]
\centerline{\includegraphics[width=10cm]{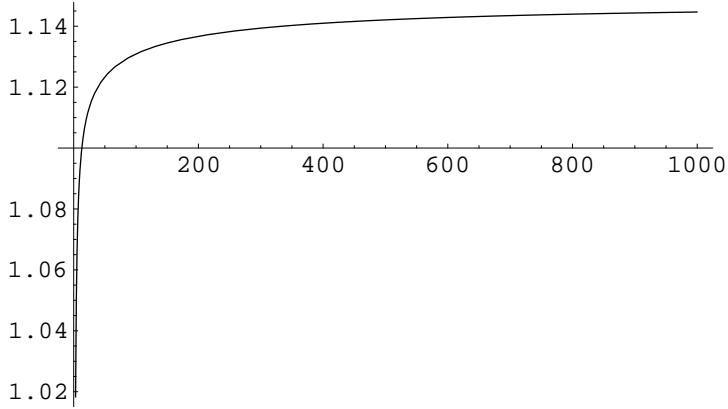}}
\caption{\label{fig:fymm2} The anomalous moment of the electron in
units of the Schwinger term (${\alpha\over 2 \pi}$) plotted versus
the PV photon mass, $\mu_1$.}
\end{figure}

Using the double-pole  prescription,
the normalization condition is quite  complicated.
An approximation sufficient for our needs is
\begin{eqnarray}
    N^2 &=& 1 + \frac{\alpha}{2\pi}\Biggl(
  \frac{1}{4} \Bigl(1 + 12\,{m_0^2} - 4\,{m_0^4} +
          2\,{m_0^4}\,\bigl( -7 + 2\,{m_0^2} \bigr) \,
          \log (\frac{m_0^2}{1 - {m_0^2}}) \\
  &-& 2 \,\log (1 - {m_0^2}) \Bigr)  + {\frac{-1 + 7\,
                               {m_0^2}}{3\,{{\mu_1}^2}}} +
   {\frac{\log ({{\mu_1}^2})}{2}} \Biggr) . \nonumber \label{n2}
\end{eqnarray}
We now have the wave function, the bare mass, $m_0$, and the wave
function normalization as functions of $\mu_1$.  We can therefore
calculate the anomalous moment using the wave function overlap formula
of~\cite{Brodsky:1980zm}.  We find
\begin{equation}
    a_e = {\alpha \over \pi^2 N^2}\int dx\; d^2k_\perp {m_0\over x}
\left[{1\over (1 - {m_0^2 + k_\perp^2 \over x} - { k_\perp^2 \over 1-x})^2}
                   - {1\over (1 - {m_0^2 +
k_\perp^2 \over x} - {\mu_1^2 + k_\perp^2 \over 1-x})^2}\right] .
\end{equation}
Here again we encounter a double pole in an integrand, and we use the
same prescription as above.  With that prescription, and making the
approximation in the second term that $m_0 = 1$, which is good
enough for our needs, we find that in units of the Schwinger term,
${\alpha \over 2 \pi}$, the anomalous moment is
\begin{eqnarray}
   a_e &=& m_0 \Bigl(-1 + 2 m_0^2 + 2 m_0^2 (1-m_0^2)
                         \ln[\frac{m_0^2}{1 - m_0^2}] \\
   &-& ({\frac{8 - 6\,{{\mu_1}^2} - 3\,{{\mu_1}^4} + 2\,{{\mu_1}^6} -
        6\,{{\left( -1 + {{\mu_1}^2} \right) }^2}\,\log (-1 + {{\mu_1}^2})}{
      3\,{{\left( -2 + {{\mu_1}^2} \right) }^4}}}) \Bigr)\frac{1}{N^2} .
      \nonumber
\end{eqnarray}
For values of $\mu_1$ larger than about 10 this is well approximated by
\begin{equation}
   a_e = m_0 \Bigl(-1 + 2 m_0^2 + 2 m_0^2 (1-m^2) \ln[\frac{m_0^2}{1 - m_0^2}]
   - \frac{2}{3 \mu_1^2} \Bigr)\frac{1}{N^2} .
\end{equation}
With this expression we find that the anomalous moment  $a_e$ is
1.02 at $\mu_1 = 3$, 1.09 at $\mu_1 = 10$, 1.13 at $\mu_1 = 100$ and
1.14 at $\mu_1 = 1000$.  We show a plot of the function in
Fig.~\ref{fig:fymm2}.

Should we view these results as satisfactory or not and what should
we choose for $\mu_1$?  In the subspace which we have kept, we
represent all the processes that contribute to the Schwinger term.
We  also include  processes  which contribute to higher order
including all orders.  If all the contributions were positive we
might therefore expect to do better than the Schwinger term, but
that is not the case: there is a considerable cancellation among the
terms which contribute to the next order contribution, the
Sommerfield--Petermann order $\alpha^2$ term~\cite{SommerfieldPetermann}.
Thus the best we can expect at this level of truncation
is to get an answer close to the Schwinger term.  If we
choose a value of $\mu_1$ anywhere between 3 times the electron mass
and 1000 times the electron mass, we obtain an answer within
15\% of the Schwinger term.
We consider that to be entirely satisfactory.

In order to estimate  an optimal value for $\mu_1$,
we apply the procedure suggested in the
Introduction:  we estimate the error
associated with having $\mu_1$ too small as $ \sim{m^2_e\over \mu^2_1}$.
We estimate the error associated with
having the PV mass too large by performing a first-order
perturbation calculation using the projection of $P^-$ onto
all of the excluded sectors of the representation space as the
perturbing operator.  The calculation proved to be quite challenging.  
The numerical effort to get even an approximate value for the magnitude of 
the projection of the perturbed wave function onto the three-body sector was 
greater than that required for the nonperturbative calculation of the magnetic 
moment.  We have done the calculation for two values of $\mu_1$, and we find 
that
\begin{equation}
   |\langle \Phi_{+}^\prime|\Phi_{+}^\prime\rangle|_{\mu_1 = 3} 
        \simeq 2 \times 10^5 \left({\alpha \over 4 \pi}\right)^2 ,
\end{equation}
\begin{equation}
   |\langle \Phi_{+}^\prime|\Phi_{+}^\prime\rangle|_{\mu_1 = 100} 
         \simeq 8 \times 10^5 \left({\alpha \over 4 \pi}\right)^2 .
\end{equation}
We can interpolate between these two values using either linear or logarithmic 
interpolation (asymptotically it should be logarithmic, but we do not know if 
we are in that region); for the present case there is little difference since 
the result is so near $\mu_1 = 3$.  Setting the two types of error equal 
to each other, we estimate the optimum value for $\mu_1$ to be between 3.5 
and 4.  This gives us an estimate of the electron's magnetic moment of about 
1.02.  There are unknown factors on both sides of the relation we used to 
estimate the optimum value of $\mu_1$, so a considerable uncertainty must 
be attached to our result.  It is interesting that the estimate suggests
that we should use a value near the lower end of the range we show in 
Fig.~\ref{fig:fymm2}.  In the present case that is the region where our 
result is the best; but we do not know if this is fortuitous.  We feel 
that the main points are that the estimate is reasonable and the final 
result is not particularly sensitive to the choice we make, even if we 
change it by an order of magnitude or more.

\section{Light-Cone Gauge}
\label{sec:lc}
In this section the notation that we use for light-cone coordinates is
\begin{equation}
x^\pm = {1\over \sqrt{2}}(x^0 \pm x^3)\,,\;\; \vec{x}_\perp=(x^1,x^2).
\end{equation}
The time coordinate is $x^+$, and the dot product of two
four-vectors is
\begin{equation}
p\cdot x=(p^+x^- + p^-x^+)
                -\vec{p}_\perp\cdot\vec{x}_\perp.
\end{equation}
The $\gamma$ matrices are chosen as follows:
\begin{eqnarray}
&\g^0=\ls
\begin{array}{cc}
0&-I\\
I&0
\end{array}
\rs,\qquad &\g^+=i\sqrt{2}\ls
\begin{array}{cc}
0&0\\
I&0
\end{array}
\rs,\nonumber\\
&\g^-=i\sqrt{2}\ls
\begin{array}{cc}
0&-I\\
0&0
\end{array}
\rs,\qquad &\g^k=i\ls
\begin{array}{cc}
-\si_k&0\\
0&\si_k
\end{array}
\rs .
\end{eqnarray}
Now we shall again attempt the calculation of the magnetic moment in
light-cone gauge,
but this time by applying the method of regularization and renormalization that
gives a light-cone perturbation theory equivalent to the standard,
dimensionally regularized Feynman series to all orders in the coupling
constant~\cite{Paston:2000fq}. This method includes the introduction of
higher derivatives into the Lagrangian, along with the use of three PV
electrons and one ``technical'' photon. Certain cutoffs, discussed below, are
also required.  From the perturbative equivalence with Feynman methods,
it is clear that the method preserves gauge invariance
in the limit where regularization is removed, at least at the level
of perturbation theory.  The Fock-space truncation will break gauge invariance,
as it does in the Feynman gauge.  The differences between the results in this
section and those in the previous section give some measure of the effect of
the breaking of gauge invariance by the truncation.

The starting form of the Lagrangian density is
\begin{eqnarray}
L=&&-\frac{1}{4}\sum_{j=0,1}(-1)^j F_{\m\n,j}\ls
1-\frac{\dd_{\p}^2}{\La^2}+\frac{2\dd_+\dd_-}{\La^2}
\ls 1+\frac{\La^2}{\m^2}\rs^j\rs F_j^{\m\n}\nonumber\\
&&+\sum_{l=0}^3 \frac{1}{\n_l}\bar\ps'_l\ls i\g^\m\dd_\m-m_l\rs\ps'_l+
e\bar\ps'\g^\m A_\m\ps',
\end{eqnarray}
where $F_{\m\n,j}=\dd_\m A_\n-\dd_\n A_\m$,
$A_\mu=A_{\m,0}+A_{\m,1}$, $A_{-,j} =0$, $\ps'=\sum_{l=0}^3\ps'_l$,
$\sum_{l=0}^3\n_l=\sum_{l=0}^3\n_lm_l=\sum_{l=0}^3\n_lm_l^2=0$,
$\n_0=1$, and $\dd_{\pm}=(\dd_0\pm\dd_3)/\sqrt{2}$.
We should comment on the photon fields labelled with $j = 1$, which we
call the technical fields.  In~\cite{Paston:2000fq}, it was explained that
these fields implement the Mandelstam--Leibbrandt (ML) prescription for the 
spurious singularity.  In the nonabelian theory, it is known that this 
prescription is necessary for a consistent perturbation theory.  In the 
abelian theory, it is not known whether the ML prescription is necessary or 
if, for instance, the principal-value prescription will suffice~\cite{bas}.  
On that basis, one might wonder whether the $j = 1$ fields are needed in QED.  
However, the $j=1$ fields are necessary for the proof in~\cite{Paston:2000fq} 
of the equivalence between light-cone and covariant Feynman formulations of
perturbation theory in either the abelian or the nonabelian case.  What we 
find in the present calculations is that the $j = 1$ fields are necessary 
even for this relatively simple problem in QED.

For the UV regularization of the photon field, we use
higher derivatives, which do not break gauge invariance in QED.
To construct a Hamiltonian formalism on the light-cone, we rewrite 
the system in a form which includes
only first derivatives of the fields with respect to $x^+$.
To regulate the theory in a way which guarantees the equivalence 
between light-cone and Feynman perturbation theory,
we must also introduce PV electrons with flavor-changing currents.  
This breaks gauge invariance, as it does in the Feynman gauge.
However, as in Feynman gauge, that breaking of gauge invariance does
not require counterterms, since we can take the limit 
$m_{1,2,3}\to \infty$. The truncation of the Fock space excludes
fermion pair creation, so there are no fermion loops;
therefore, after taking 
the limit $\ep\to 0$ (i.e. removing the light-cone
regularization $|k_-|\ge\ep>0$),
we can take the limit $m_{1,2,3}\to \infty$ at fixed $\La$.  
We then find that gauge invariance is restored without using  
counterterms in the Lagrangian.
The calculations that we carry out here confirm  that
these limits indeed turn out to be finite.
If we had included enough of the representation space
to allow pair production to occur, then we would not be able to take the limit
$m_{1,2,3}\to \infty$.  In that case, we would have to add counterterms
and also renormalize the electric charge $e$.

To get a canonical light-cone formulation and determine the Hamiltonian $P^-$, 
we decompose the bi-spinors $\ps'$ into two-spinor components,
$\ps'_l=\{\ps_{l,+},\ps_{l,-}\}$, and express the components
$\ps_{l,-}$ in terms of other field variables using the canonical
constraint  equation on the light-cone. As shown in~\cite{Paston:2000fq},
it is convenient to use $\f_j\equiv \dd_\m A^\m_j$ and $\f\equiv \dd_\m A^\m$
in place of the $A_{+,j}$ as degrees of freedom, and we shall make that choice.
With all these choices, $P^-$ can be written as
\begin{eqnarray}
&&  P^-=\int dx^-\int d^2x^\p
\sum_{j=0,1}\Biggl\{\frac{\La^2(-1)^{j+1}}{8\ls
1+\frac{\La^2}{\m^2}\rs^j}
\nonumber\\
&&\times\sum_{k=1,2}\ls (-1)^j\dd_-^{-1}\Pi_{k,j}-\ls
1-\frac{\dd_\p^2}{\La^2}\rs A_{k,j}+ \frac{\dd_\p^2}{\La^2}\ls
1+\frac{\La^2}{\m^2}\rs^j A_{k,j}\rs^2 \nonumber\\&&
+\frac{(-1)^{j+1}}{2}\sum_{k=1,2}A_{k,j}\dd_\p^2 \ls
1-\frac{\dd_\p^2}{\La^2}\rs A_{k,j}+ \frac{(-1)^{j+1}}{2}\f_j \ls
1-\frac{\dd_\p^2}{\La^2}\rs \f_j\Biggr\} \nonumber\\&&
+\frac{i}{\sqrt{2}}\sum_{l=0}^3\n_l^{-1}\ps_{l,+}^+(\dd_\p^2-m_l^2)
\dd_-^{-1}\ps_{l,+}-
e\sqrt{2}\ps_+^+\ps_+\dd_-^{-1}\ls\f+\sum_{k=1,2}\dd_k A_k\rs
\nonumber\\
&& +\frac{e}{\sqrt{2}}\ls\ps_+^+ \hat A_\p \dd_-^{-1}
\sum_{l=0}^3(\dd_\p^2-m_l^2)\ps_{l,+}+h.c.\rs, \label{eq:pmlc}
\end{eqnarray}
where $\Pi_{k,j}=\frac{\de L}{\de(\dd_+ A_{k,j})}$ are canonical
momenta conjugate to  $A_{k,j}$, and we have defined
$\hat A_{\p}=A_1\si_1+A_2\si_2$ and $\dd_{\p}=(\dd_1,\dd_2) $.

We decompose our field variables in terms of creation and
annihilation operators acting in light-cone Fock space:
\begin{equation}
A_{k,j}(x)=(2\pi)^{-3/2}\int d^3p\sum_{\z=0,1} \frac{a_{j\,\z
k}(p)e^{-ipx}+h.c.}{\sqrt{2p^+\ls\frac{p_\p^2}{\m^2}-1\rs^j}},
\end{equation}
\begin{equation}
\Pi_{k,j}(x)=-i(2\pi)^{-3/2}\int d^3p\sum_{\z=0,1}
\sqrt{\frac{p^+}{2}\ls\frac{p_\p^2}{\m^2}-1\rs^j} \ls a_{j\,\z
k}(p)e^{-ipx}-h.c. \rs,
\end{equation}
\begin{equation}
\f_j(x)=i(2\pi)^{-3/2}\La\ls 1+\frac{\La^2}{\m^2}\rs^{-j/2}\int d^3p
\frac{a_{j\,\z k}(p)e^{-ipx}-h.c.}{\sqrt{2p^+}},
\end{equation}
\begin{equation}
\ps_{l,+,s}(x)=2^{-1/4}(2\pi)^{-3/2}\int d^3p\ls
b_{l,s}(p)e^{-ipx}+d^+_{l,s}(p)e^{ipx}\rs,
\end{equation}
where $s$ is a spin projection.  The integrals are defined as
\begin{equation}
\int d^3p\equiv \int_{p_\p^2>v^2}d^2p_\p\int_\ep^\infty dp^+,
\end{equation}
where $v$ and $\ep>0$ are regulation parameters which must be taken to
zero in a prescribed order, as was shown in~\cite{Paston:2000fq}, and will be
discussed below.
The nonzero commutation relations for the creation and annihilation operators
have the following form:
\begin{eqnarray}
&& [a_{j\,\z k}(p),a_{j'\,\z'
k'}^+(p')]=(-1)^\z\de_{kk'}\de_{jj'}\de_{\z\z'} \de^3(p-p'),
\nonumber\\&&
[a_{j}(p),a_{j'}^+(p')]=(-1)^{j+1}\de_{jj'}\de^3(p-p'),
\\&&
[b_{l,s}(p),b_{l',s'}^+(p')]= [d_{l,s}(p),d_{l',s'}^+(p')]=
\n_l\de_{ll'}\de_{ss'}\de^3(p-p').\nonumber
\end{eqnarray}

Substituting these decompositions into (\ref{eq:pmlc}), and keeping only the
terms necessary for our approximate calculation of the anomalous magnetic 
moment, we obtain
\begin{eqnarray}
&& P^- =\sum_{l,s}\int d^3p\; \n_l^{-1} E_l(p) b_{l,s}^+(p)b_{l,s}(p)+
\sum_{j}\int d^3k (-1)^{j+1} E_j(k) a_{j}^+(k)a_{j}(k)
\nonumber\\&&
+\sum_{j,\,\z,\la}\int d^3k (-1)^{\z} E_{j\,\z}(k)
a_{j\,\z\la}^+(k) a_{j\,\z\la}(k)+
\frac{e}{(2\pi)^{3/2}}\sum_{l,l',j,s}\int d^3q\int d^3k
\nonumber\\&&
\qquad\times\ls a_j(k)b_{l,s}^+(q+k)b_{l',s}(q)+
a_j^+(k)b_{l,s}^+(q-k)b_{l',s}(q)\rs \frac{\La}{\ls
1+\frac{\La^2}{\m^2}\rs^{j/2}\sqrt{2{k^+}^3}} \nonumber\\&&
+\sum_{\z,\la}\frac{1}{\sqrt{2k^+\ls\frac{k_\p^2}{\m^2}-1\rs^j}}
\Biggr[ a_{j\,\z\la}(k)b_{l,s}^+(q+k)b_{l',s}(q) \nonumber\\&&
\qquad\qquad\times\ls\frac{q_{(-\la)}}{q^+}\de_{s,-\la/2}+
\frac{(q+k)_{(-\la)}}{(q+k)^+}\de_{s,\la/2}-\frac{k_{(-\la)}}{k^+}\rs
\nonumber\\&&
\qquad +a_{j\,\z\la}^+(k)b_{l,s}^+(q-k)b_{l',s}(q)
\ls\frac{q_{(\la)}}{q^+}\de_{s,\la/2}+
\frac{(q-k)_{(\la)}}{(q-k)^+}\de_{s,-\la/2}-\frac{k_{(\la)}}{k^+}\rs
\nonumber\\&&
\qquad +\frac{i}{\sqrt{2}}\ls
\frac{m_l}{(q+k)^+}-\frac{m_{l'}}{q^+}\rs
a_{j\,\z\la}(k)b_{l,s}^+(q+k)b_{l',-s}(q)\de_{s,-\la/2}
\nonumber\\&&
\qquad +\frac{i}{\sqrt{2}}\ls
\frac{m_l}{(q-k)^+}-\frac{m_{l'}}{q^+}\rs
a_{j\,\z\la}^+(k)b_{l,s}^+(q-k)b_{l',-s}(q)\de_{s,\la/2}\Biggr],
\label{lc1}
\end{eqnarray}
where we use helicity components for the operators $a_{j\,\z k}(p)$:
\begin{equation}
a_{j\,\z \la}\equiv \frac{1}{\sqrt{2}}\ls a_{j\,\z 1}+i\la a_{j\,\z
2}\rs, \qquad \la=\pm 1,
\end{equation}
and we have defined
\begin{eqnarray}
&& p_{(\la)}\equiv \frac{1}{\sqrt{2}}\ls p_1+i\la p_2\rs,\quad
p_\p^2\equiv p_1^2+p_2^2=\sum_{\la=\pm
1}p_{(\la)}p_{(-\la)}, \nonumber\\&& E_l(p)\equiv
\frac{p_\p^2+m_l^2}{2p^+}, E_j(k)\equiv \frac{k_\p^2+\La^2}{2k^+\ls
1+\frac{\La^2}{\m^2}\rs^j}, E_{j\,\z}(k)\equiv
\frac{k_\p^2+\z\La^2}{2k^+\ls 1+\z\frac{\La^2}{\m^2}\rs^j} .
\end{eqnarray}

Now we truncate the Fock space and write the wave function as
\begin{eqnarray}
|p\rangle=&&\Biggl(\sum_{l,s}f_l^s(p)b_{l,s}^+(p)+ \sum_{l,j,s}\int
d^3 q\; f_{lj}^s(p,q)b_{l,s}^+(q)a_j^+(p-q)
\nonumber\\&&
+\sum_{l,j,\z,\la,s}\int d^3 q\; f_{lj\,\z\la}^s(p,q)b_{l,s}^+(q)
a_{j\,\z\la}^+(p-q) \Biggr)|0\rangle. \label{lc2}
\end{eqnarray}
As in the previous section, we can choose a state with $p_{\p}=0$.  The
eigenvalue problem in this subspace is
\begin{equation}
    P^+ P^-|p\rangle=\frac{M^2}{2}|p\rangle. \label{lc3}
\end{equation}

Acting with $P^-$, as given by (\ref{lc1}), on the state (\ref{lc2}), 
Eq.~(\ref{lc3}) gives the following relations among the coefficients of 
the basic state vectors.  For $b_{l,s}^+(p)|0\rangle$ we have
\begin{eqnarray}
&& \frac{m_l^2-M^2}{2}f_l^s(p)+\frac{e}{(2\pi)^{3/2}}\sum_{l',j}
\int d^3 q\; f_{l'j}^s(p,q)
\frac{\n_{l'}(-1)^{j+1}\La}{\sqrt{2(1-x)^3}\ls
1+\frac{\La^2}{\m^2}\rs^{j/2}} \nonumber\\&&
+\sum_{l',j,\z,\la,s'}\int d^3 q\; f_{l'j\,\z\la}^{s'}(p,q)
\frac{\n_{l'}(-1)^{\z}}{\sqrt{2(1-x)}\ls \frac{z}{\m^2}-1\rs^{j/2}}
\Biggl[ \de_{ss'}\Biggl( \de_{s,\la/2}p_{(-\la)} \nonumber\\&&
+\de_{s,-\la/2}
\frac{q_{(-\la)}}{x}-\frac{(p-q)_{(-\la)}}{1-x}\Biggr)+
\de_{s',-s}\frac{i}{\sqrt{2}}\ls
m_l-\frac{m_{l'}}{x}\rs\de_{s,-\la/2}\Biggr]=0, \label{lc4}
\end{eqnarray}
where $x\equiv q^+$, $z\equiv (p-q)_\p^2$.
For $b_{l,s}^+(q)a_j^+(p-q)|0\rangle$ there is
\begin{eqnarray}
&& f_{lj}^s(p,q)\ls E_l(q)+E_j(p-q)-E(p)\rs \nonumber\\&&
\qquad +\frac{e}{(2\pi)^{3/2}}\sum_{l'}
\frac{\n_{l'}f_{l'}^s(p)\La}{\sqrt{2(1-x)^3\ls
1+\frac{\La^2}{\m^2}\rs^j}}=0. \label{lc4.1}
\end{eqnarray}
For $b_{l,s}^+(q)a_{j\,\z\la}^+(p-q)|0\rangle$ we find
\begin{eqnarray}
&& f_{lj\,\z\la}^s(p,q)\ls E_l(q)+E_{j\,\z}(p-q)-E(p)\rs
\nonumber\\&&
+ \frac{e}{(2\pi)^{3/2}}\sum_{l',s'}
\frac{\n_{l'}f_{l'}^{s'}(p)}{\sqrt{2(1-x)\ls \frac{z}{\m^2}-1\rs^j}}
\Biggl[ \de_{ss'}\Biggl( \de_{s,\la/2}p_{(\la)} \nonumber\\&&
\qquad  +\de_{s,-\la/2}
\frac{q_{(\la)}}{x}-\frac{(p-q)_{(\la)}}{1-x}\Biggr)+
\de_{s',-s}\frac{i}{\sqrt{2}}\ls
\frac{m_{l}}{x}-m_{l'}\rs\de_{s,\la/2}\Biggr]=0. \label{lc4.2}
\end{eqnarray}
Expressing the wave functions, $f_{lj}^s(p,q)$ and $f_{lj\,\z\la}^s(p,q)$,
in terms of the $f_l^s(p)$ and substituting into the relation
(\ref{lc4}), we obtain the eigenvalue equation
\begin{eqnarray}
&& \ls m_l^2-M^2\rs f_l^s= \frac{\al}{\pi}
\sum_{l_2=0}^3f_{l_2}^s \n_{l_2}
\int\limits_{v^2}^{\infty}dz\int\limits_{\ep}^{1-\ep}dx
\sum_{l_1=0}^3\n_{l_1} \nonumber\\&&
\times \sum_{j=0,1} \left\{
\frac{(-1)^{j+1}\La^2}{\ls 1+\frac{\La^2}{\mu^2}\rs^j(1-x)^3 \ls
\frac{z+m_{l_1}^2}{x}+\frac{z+\La^2}{(1-x)\ls
1+\frac{\La^2}{\mu^2}\rs^j}-M^2\rs}\right.  \nonumber\\&&
+\sum_{\z=0,1}\frac{(-1)^\z}{2\ls\frac{z}{\mu^2}-1\rs^j(1-x)} \left[
\frac{z(1+x^2)}{x^2(1-x)^2}+\ls m_l-\frac{m_{l_1}}{x}\rs\ls
m_{l_2}-\frac{m_{l_1}}{x}\rs \right] \nonumber\\&&
\qquad\qquad\times\left. \frac{1}{\ls
\frac{z+m_{l_1}^2}{x}+\frac{z+\z\La^2}{(1-x)\ls
1+\z\frac{\La^2}{\mu^2}\rs^j} -M^2\rs}\right\}, \label{lc4.3}
\end{eqnarray}
where we have defined $z=q_{\p}^2$.

We write this equation in the form
\begin{equation}
m_0^2-M^2={{\al}\over{\pi}} \bigl(
AM^2+2Bm_0+C\bigr)+{{\al^2}\over{\pi^2}} \bigl(A
C-B^2\bigr), \label{lc5}
\end{equation}
where $A$, $B$, and $C$ are real functions of $m_0$, $M$, and the
regularization parameter $\La$.  To restore Lorentz invariance and
gauge invariance, except for the breaking due to the truncation, we must
now take the following limits in the order given~\cite{Paston:2000fq}:
\begin{equation}
\ep\to 0,\quad \mu\to 0,\quad v\to 0,\quad m_1,m_2,m_3\to\infty,
\label{lc6}
\end{equation}

We can now find the value of $m_0$ by setting $M$ equal to 1 and solving
(\ref{lc5}) numerically.  We can find the leading order solution by expanding
the expressions for $A$, $B$, and $C$ for large values of  $\La$.  We find
\begin{equation}
A=\frac{1}{4}\ln\La^2+\dots,\quad
B=-\frac{m_0}{2}\ln\La^2+\dots,\quad
C=-\frac{3M^2}{4}\ln\La^2+\dots.
\end{equation}
Keeping only these terms, we can solve Eq.~(\ref{lc5}) and obtain
\begin{equation}
m_0^2=M^2
\frac{1-\frac{1}{2}\ls\frac{\al}{\pi}\ln\La^2\rs
             -\frac{3}{16}\ls\frac{\al}{\pi}\ln\La^2\rs^2}%
{1+\ls\frac{\al}{\pi}\ln\La^2\rs
         +\frac{1}{4}\ls\frac{\al}{\pi}\ln\La^2\rs^2}.
\end{equation}
This leads to the following expression in the lowest order in
coupling constant:
\begin{equation}
m_0^2=M^2\ls 1-\frac{3\al}{2\pi}\ln\La^2\rs .
\end{equation}
This expression agrees with Eq.~(\ref{eq:m0f}) (and with the known one-loop
result for the self-energy of the electron).  As expected in QED,
$m_0^2<M^2$. Therefore, the integrand of equation
(\ref{lc4.3}) does contain a simple pole that we have integrated
using the principal-value prescription, as we proposed in the Introduction.

For either value of $s$, we can parameterize the constants, $f_l^s$, in
the wave function in terms of $A$ and $B$ as
\begin{equation}
f_l^s= \Biggl\{{{1+{{\al}\over{\pi}}A}\over m_0 A+B},\quad
{{\al}\over{\pi}}{1\over m_1},\quad {{\al}\over{\pi}}{1\over
m_2},\quad {{\al}\over{\pi}}{1\over m_3}\Biggr\} .
\label{lc7}
\end{equation}
To obtain a normalized wave function we must divide by the norm $N$,
which is given by the expression
\begin{eqnarray}
&& N^2=\sum_{l=0}^3\n_l|f_l|^2+
\left|\sum_{l_1=0}^3\n_{l_1}f_{l_1}\right|^2\frac{\al}{\pi}
\nonumber\\&&
\times\Biggl(\;
\int\limits_{v^2}^{\infty}dz\int\limits_{\e}^{1-\e}dx
\sum_{l=0}^3\sum_{j=0,1}
\frac{(-1)^{j+1}\n_l\La^2}{(1-x)^3\ls 1+\frac{\La^2}{\mu^2}\rs^j%
\ls\frac{z+m_l^2}{x}+\frac{z+\La^2}{(1-x)\ls 1+\frac{\La^2}{\mu^2}\rs^j}%
-M^2\rs^2} \nonumber\\&&
+\int\limits_{v^2}^{\infty}dz\int\limits_{\e}^{1-\e}dx
\sum_{l=0}^3\sum_{j=0,1}\sum_{\z=0,1}
\frac{(-1)^\z \n_l z(1+x^2)}{2(1-x)^3x^2\ls \frac{z}{\mu^2}-1\rs^j%
\ls\frac{z+m_l^2}{x}+\frac{z+\z\La^2}{(1-x)\ls 1+\z\frac{\La^2}{\mu^2}\rs^j}%
-M^2\rs^2}\;\Biggr) \nonumber\\&&
+\sum_{l=0}^3\left|\sum_{l_1=0}^3\n_{l_1}f_{l_1}\ls\frac{m_l}{x}-m_{l_1}\rs
\right|^2\frac{\al}{2\pi} \nonumber\\&&
\times\int\limits_{v^2}^{\infty}dz\int\limits_{\e}^{1-\e}dx
\sum_{j=0,1}\sum_{\z=0,1}
\frac{(-1)^\z \n_l}{(1-x)\ls\frac{z}{\mu^2}-1\rs^j%
\biggl(\frac{z+m_l^2}{x}+\frac{z+\z\La^2}{(1-x)\ls
         1+\z\frac{\La^2}{\mu^2}\rs^j}%
-M^2\biggr)^2}. \label{lc8}
\end{eqnarray}
Here again we encounter the double pole, which we treat according to the
prescription (\ref{prescr}) (see the Introduction).

The result of calculating the integrals in (\ref{lc8}), and taking the proper
limits specified in (\ref{lc6}), is a very complicated expression
that is a function of $m_0$, $M$, and $\La$.  The limits must be taken only
after performing the integrals of Eq.~(\ref{lc8}).

To calculate the anomalous magnetic moment of the electron, we need to use
the norm of the eigenstate (\ref{lc8}) and calculate the matrix
element of the current component $j^+(x)$ between the states (\ref{lc2}).
Using the method of~\cite{Brodsky:1980zm}, we can write the expression
for the anomalous magnetic moment $a_e$ as
\begin{equation}
a_e=-\frac{2M}{N^2}
i\frac{\dd}{\dd p_1} \langle
p_\perp = p_1;s|j^+(0)|p_\perp = 0;-s\rangle\mid_{p_1=0},
\label{lcn1}
\end{equation}
where the relevant part of the current (containing only
$b_{l,s}$ and $b_{l,s}^+$) is
\begin{equation}
j^+(0)=\sum_{l_1,l_2,s}\int d^3 q_1 d^3 q_2\,
b_{l_1,s}^+(q_1)b_{l_2,s}(q_2).
\label{lcn2}
\end{equation}
Using the expression in (\ref{lc2}) and Eqs.~(\ref{lc4.1}) and
(\ref{lc4.2}), we get
\begin{eqnarray}
& a_e&=
\frac{M e^2}{2(2\pi)^3N^2}
\sum_{l_1,l_2,l_3,l_4,j,\z}
(-1)^\z
\n_{l_1}\n_{l_2}\n_{l_3}\n_{l_4}f_{l_3}f_{l_4}^* \nonumber\\&&
\times\int d^3 q
\frac{1}{(1-x)\ls
\frac{z}{\m^2}-1\rs^j} \ls
\frac{m_{l_1}}{x}-m_{l_3}\rs
\nonumber\\&&
\times\ls E_{l_1}(q)+E_{j\,\z}(p-q)-\frac{M^2}{2}\rs^{-1}
\ls E_{l_2}(q)+E_{j\,\z}(p-q)-\frac{M^2}{2}\rs^{-1}.
\label{lcn3}
\end{eqnarray}
In the calculation of this integral, we have again used
the prescription (\ref{prescr}) for the double pole.

One can show that after removing the regularization,
(\ref{lc6}), we have  $f^s_1=f^s_2=f^s_3=f^s_{l,0}=0$.
Then the calculation gives an expression which can be written at
large $\La$ as
\begin{eqnarray}
&& a_e=
\frac{\al f_0}{2\pi M^5N^2}
\Biggl(
f_0M^2m_0\ls 2m_0^2-M^2\rs-\frac{\al}{\pi}M^2\ls 2m_0^2+M^2\rs
\nonumber\\&&\qquad\qquad\qquad\qquad\qquad
-2m_0^3\ls f_0\ls M^2-m_0^2\rs+\frac{\al}{\pi}m_0\rs
\log\left|1-\frac{M^2}{m_0^2}\right|\Biggr)\nonumber\\&&\qquad
-\frac{\al f_0 M}{6\pi N^2}
\ls 2m_0f_0+\frac{\al}{\pi}\rs\frac{1}{\La^2}+
O\ls\frac{1}{\La^4}\rs.
\label{lcn4}
\end{eqnarray}
Substituting the values of $m_0$, $f_0$, and $N$, found from
(\ref{lc5}), (\ref{lc7}), and (\ref{lc8}) at fixed $\al$ and $\La$,
and setting $M=1$, we get the value of $a_e$.
At lowest order in $\al$ (when $m_0\to M$, $N\to f_0$), the
expression (\ref{lcn4})
gives the Schwinger term $a_e=\frac{\al}{2\pi}$.

\begin{figure}[bhtp]
\hskip -7mm\centerline{\includegraphics[width=9cm]{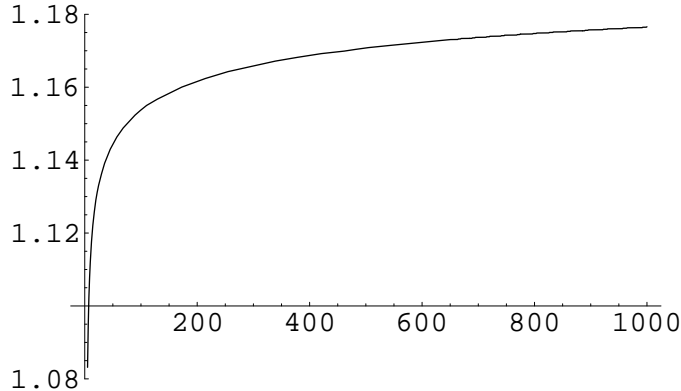}}
\caption{\label{fig:lcmm}
The result of a light-cone calculation of the anomalous moment of the
electron in units of the Schwinger term (${\alpha\over 2 \pi}$) plotted
versus the UV regularization parameter $\La$.}
\end{figure}

In Fig.~\ref{fig:lcmm} we show the value of
the anomalous magnetic moment, $a_e$, in units of the
Schwinger term $\ls \frac{\al}{2\pi}\rs$, plotted against the UV
regularization parameter $\La$.

\section{Discussion}
\label{sec:Discussion}

In this paper we have presented nonperturbative calculations of the 
electron's magnetic moment.  In each case we used the light-cone 
representation and solved the eigenvalue equation in a basis truncated 
to include only states with one electron and states with one electron 
and one photon.  Once we found the one-electron eigenstate in the 
truncated subspace, we then calculated the magnetic moment from the eigenstate.

We found that there were three problems that we had to overcome in 
order to make a successful calculation: the problem of uncancelled 
divergences, the problem of new divergences, and the problem of 
maintaining gauge invariance.  Our solution to the problem of uncancelled 
divergences is to keep the PV photon mass (or the, nearly equivalent, 
UV regulator $\La$ in the light-cone gauge case) finite.  For that 
method to succeed we had to find a large range of PV photon masses for 
which the error due to including the negative metric states 
in our wave function and the error due to the truncation of the 
representation space were both small.  We did find such a range.  
If one accepts the errors shown in Figs.~\ref{fig:fymm2} and 
\ref{fig:lcmm} as being not so large as to render the calculation useless, 
we must only estimate the optimum PV mass as lying between three 
electron masses and one thousand electron masses in order to perform a 
useful calculation.  The estimate we obtained, $\approx 4 m_e$, not only
lies in this range but, perhaps fortuitously, lies in the part of the range
in which our answer is nearest to the correct answer.

In our calculations we encounter integrals whose integrands contain poles, 
which we define as principal values, and double poles.  In the case of the 
double pole we have provided a prescription which is the nearest we can come 
to defining the double pole as the derivative of a principal value.  With 
these prescriptions we obtain a successful calculation.  The ultimate test 
of the prescriptions is whether or not they respect the Ward identity.  We 
have not proven that the prescriptions preserve the Ward identity beyond 
the present calculations.  An unexpected feature of the prescriptions is 
that the wave function normalization is now infrared finite, whereas it is 
infrared divergent in perturbation theory.  We have therefore been able to 
carry out the present calculations using zero for the mass of the physical 
photon.

The most difficult problem we had to solve was the problem of maintaining 
gauge invariance.  If the standard light-cone procedures are applied, 
regulated with any combination of PV fields that we tried (or, regulated 
with a momentum cutoff), the result is not a successful calculation.  The 
failure of the standard calculation to produce useful results can be traced 
to the loss of gauge invariance.  Understanding the details of that loss of 
gauge invariance is an interesting unsolved problem.  We have produced 
successful calculations by the use of Feynman gauge and by the use of 
light-cone gauge regulated with both higher derivatives and PV fields.  
It should be noted that in either of the formulations which resulted in a 
successful calculation, $A^-$ is a degree of freedom, whereas in the 
unsuccessful light-cone gauge calculation $A^-$ satisfies a differential 
constraint equation. 

An important observation is that, with the use of the PV fields, the 
operator $P^-$ can be constructed without inverting any covariant 
derivatives.  In the past, most calculations for gauge theories in the 
light-cone representation have been done in light-cone gauge due to the 
apparent need to invert a covariant derivative in any other 
gauge.  With the use of appropriately coupled PV fields, that impediment 
to the use of other gauges, including covariant gauges, is removed.  In 
the present work the calculations in Feynman gauge are considerably simpler 
than the successful calculations in light-cone gauge.  Looking forward to 
the nonabelian case, it is not clear that this will remain the case.  
Use of the Feynman gauge would involve the inclusion of Faddeev--Popov 
fields, whereas presumably that would not be necessary in light-cone 
gauge.  Also, the nonabelian formulation corresponding to Sec.~\ref{sec:lc} 
has been shown to give perturbative equivalence to covariant methods to all 
orders in perturbation theory.  No such demonstration has yet been given 
for other gauges.

Even when we have a formulation which, in the absence of truncation, would 
preserve gauge invariance, the truncation will break gauge invariance.  We 
have argued that the question of whether or not that breaking of gauge 
invariance is acceptable is more a question of accuracy than symmetry: 
if our broken answer is close to the correct, unbroken answer without 
truncation, it will be a useful answer.  The difference between 
the answers we obtained here in the Feynman gauge and in the light-cone 
gauge provides some test of that idea.  The answers differ by some two 
percent.  We believe that the fact that they are close to each other is 
due to the fact that they are close to the answer which would result from 
a calculation without truncation.

The objective of the calculations presented here was not to improve on 
the very accurate calculations of perturbation theory.  Rather, the 
objective was to verify that the nonperturbative methods we have 
developed really do represent a nonperturbative approximation to QED.  
We have succeeded in this, not only because the answers we get 
are as close as can be expected to the correct answer but also because 
an expansion of our approximate, nonperturbative answer in powers of 
the coupling constant reproduces the finite terms of standard QED 
perturbation theory, through the order allowed by our truncation of 
the representation space.  We get additional terms up to all orders 
in the coupling constant, which can be identified as some of the terms 
of standard perturbation theory.  Therefore, it is reasonable 
to hope that our nonperturbative methods will produce useful calculations 
for problems which perturbation theory cannot address, such as hadron 
bound states.


\section*{Acknowledgments}
This work was supported by the Department of Energy through
contracts DE-AC03-76SF00515 (S.J.B.), DE-FG02-98ER41087 (J.R.H.),
and DE-FG03-95ER40908 (G.M.).



\end{document}